\documentclass[conference]{IEEEtran}
\IEEEoverridecommandlockouts
\usepackage[bottom=1.02in,top=0.71in,left=0.64in,right=0.64in]{geometry}
\usepackage{amsmath}
\usepackage{amsfonts}
\usepackage{amssymb}
\usepackage[ruled,vlined]{algorithm2e}
\usepackage{caption}
\usepackage{mathtools}  
\usepackage{amsthm}
\usepackage[]{algorithm2e}
\usepackage{booktabs}
\usepackage{subfigure}
\usepackage{tabulary}
\usepackage{footnote}
\usepackage[export]{adjustbox}
\usepackage{booktabs}
\usepackage[justification=centering]{caption}
\usepackage{comment}
\newtheorem{remark}{Remark}
\usepackage{cases}
\usepackage{graphicx}
\usepackage{cite}
\usepackage{multicol}
\usepackage{xcolor}
\usepackage{graphicx}
\graphicspath{{./}{figures/}}
\newcommand\numeq[1]%
  {\stackrel{\scriptscriptstyle(\mkern-1.5mu#1\mkern-1.5mu)}{=}}
\usepackage{stfloats}
\usepackage{cuted}
\usepackage{lipsum}  

\usepackage{dsfont}
\usepackage{amsmath,amssymb}

\newtheorem{Prob}{\textbf{Problem}}

\usepackage{optidef}
\usepackage{amsmath}
\usepackage{needspace}

\title{RIS-Aided E2E Multi-Path Uplink Transmission Optimization for 6G Time-Sensitive Services}

 \author{\text{Liu Cao}\IEEEauthorrefmark{1}, \text{Zisheng Gong}\IEEEauthorrefmark{1}, \text{Ziyue Xiao}, \text{Zhaoyu Liu}, \text{Houtianfu Wang}, \text{Lyutianyang Zhang}
\vspace{-1cm}

 \thanks{This paper has been presented in part at the IEEE International Conference on Communications in China \cite{cao2025latency}.} 
 \thanks{\IEEEauthorrefmark{1}Both authors contributed equally to this work. Liu Cao, Zisheng Gong, Ziyue Xiao, and Zhaoyu Liu are with both City University of Hong Kong (Dongguan), Dongguan, China, and City University of Hong Kong, Hong Kong (e-mail:\{liu.cao, 72404185, 72405391, 72515198\}@cityu-dg.edu.cn). 
 Houtianfu Wang is with the Department of Engineering, University of Cambridge, Cambridge, UK (email:hw680@cam.ac.uk).
 Lyutianyang Zhang is  with
the School of Microelectronics and Communication Engineering, Chongqing
University, Chongqing, China (email: zhanglyutianyang@cqu.edu.cn).  (Corresponding author: Lyutianyang Zhang.)}
\thanks{This work was supported by the Youth Innovation Talent Project of Guangdong Provincial Universities (Grant No. 2025KQNCX17).}

}

\begin{document}

\maketitle
\thispagestyle{empty}
\begin{abstract}
The Access Traffic Steering, Switching, and Splitting (ATSSS) defined in the latest  3GPP Release 19 enables traffic flow over the multiple access paths to achieve the lower-latency End-to-end (E2E) delivery for 6G time-sensitive services. However, the existing E2E multi-path operation often falls short of more stringent QoS requirements for 6G time-sensitive services. This work proposes a Reconfigurable Intelligent Surfaces (RIS)-aided E2E multi-path uplink (UL) transmission architecture that explicitly accounts for both radio link latency and N3 backhaul latency, via the coupled designs of the UL traffic-splitting ratio, transmit power, receive combining, and RIS phase shift under practical constraints to achieve the minimum average E2E latency. We develop an alternating optimization framework that updates the above target parameters to be optimized. The simulations were conducted to compare the effectiveness of the proposed E2E optimization framework that lowers the average E2E latency up to 43\% for a single user and 32\% for the whole system compared with baselines in our prior work [1].


\end{abstract}


\section{Introduction}
With the rapid growth of Internet-of-Things (IoT) devices towards 6G networks, time-sensitive mobile services have placed tremendous pressure on existing wireless networks\cite{3gpp.38.913}. Advanced technologies such as Multi-access Edge Computing (MEC), network slicing, and Ultra-Reliable Low-Latency Communications (URLLC) were leveraged to satisfy the stringent quality of service (QoS) requirements for 5G services. Among these, the End-to-end (E2E) multi-path transmission scheme has emerged as a promising solution to mitigate the E2E multi-path latency for URLLC traffic in dynamic and diverse wireless environments \cite{cao2025latency}.

Traditional cellular networks rely on single E2E path, which however struggle with handling the latency issues. While the E2E multi-path framework utilizes multiple 3GPP accesses (e.g. 5G NR Uu) or non-3GPP accesses (e.g., WLAN 802.11), within the context of Access Traffic Steering, Switching and Splitting (ATSSS) rule \cite{3gpp.38.913,3gpp.23.725}, to effectively mitigate such issues in the 5G/5G-A era \cite{cao2025latency}. The authors in \cite{ba2022multiservice} investigated an optimization problem for traffic scheduling in NR and WLAN aggregation (NWA) for enhanced mobile broadband (eMBB) services. The literature \cite{hurtig2018low} studied an E2E multi-path transmission control protocol (MPTCP) scheduler for latency-sensitive applications. A joint power control and channel allocation scheme was developed in \cite{zhao2019joint} to reduce the interference adaptively. However, 6G time-sensitive services typically are characterized with more stringent QoS requirements, meanwhile, 6G wireless environments with more complex PHY/MAC impairments fundamentally bound the achievable E2E latency mitigation, thereby limiting the effectiveness of existing E2E multi-path schemes. 

In this context, Reconfigurable Intelligent Surface (RIS) introduces a novel PHY-layer controllability by actively shaping the wireless environment \cite{wu2021intelligent,shafique2024going,zhu2024intelligent,shi2024ris}, enabling the construction of more favorable and stable E2E links. Moreover, recent space-time-coding metasurface experiments have demonstrated simultaneous independent multi-beamforming and multi-channel transmission across different spatial directions and frequencies\cite{tian2026ambient}, suggesting that a shared programmable aperture can concurrently support multiple wireless links. This additional degree of freedom motivates RIS-assisted E2E multi-path optimization, where traffic control across multiple accesses can be jointly designed with propagation-aware link enhancement, offering a viable pathway toward meeting the lower E2E multi-path latency requirements for 6G networks. 

Inspired by the aforementioned issues, we propose a RIS-assisted E2E multi-path  (UL) transmission architecture that outperforms the existing optimized E2E multi-path approaches. To the best knowledge of authors, this is the first work that integrates both benefits of the RIS and the standardized E2E multi-path architecture that well aligned with the new solution to the lower latency characteristics of 6G time-sensitive services. Compared with our preliminary conference version\cite{cao2025latency}, this paper provides a substantial extension in several aspects. First, while [1] considered a single-UE two-path E2E multi-path transmission setting, this work develops a RIS-assisted multi-BS multi-UE uplink framework for 6G time-sensitive services. Second, beyond the abstract path-level latency model in [1], this paper explicitly incorporates the RIS-assisted channels, BS receive combining, inter-user interference, and the unit-modulus RIS constraint into the E2E latency-aware design. Third, a new joint optimization of traffic splitting, transmit power, receive combining, and RIS phase shifts is formulated and solved via an alternating optimization framework. Finally, new simulation results are provided to quantify the additional latency reduction achieved by RIS over the prior E2E multi-path baselines.


The remainder of this paper is organized as follows: Section II presents the overall system architecture, Section III provides the problem formulation with the solution, Section IV details the simulation results, and Section V draws the conclusions for this paper.

\section{System Architecture}
\label{sec:sys_arc}

\begin{figure}[t]
    \centering
    \includegraphics[width=.4\textwidth]{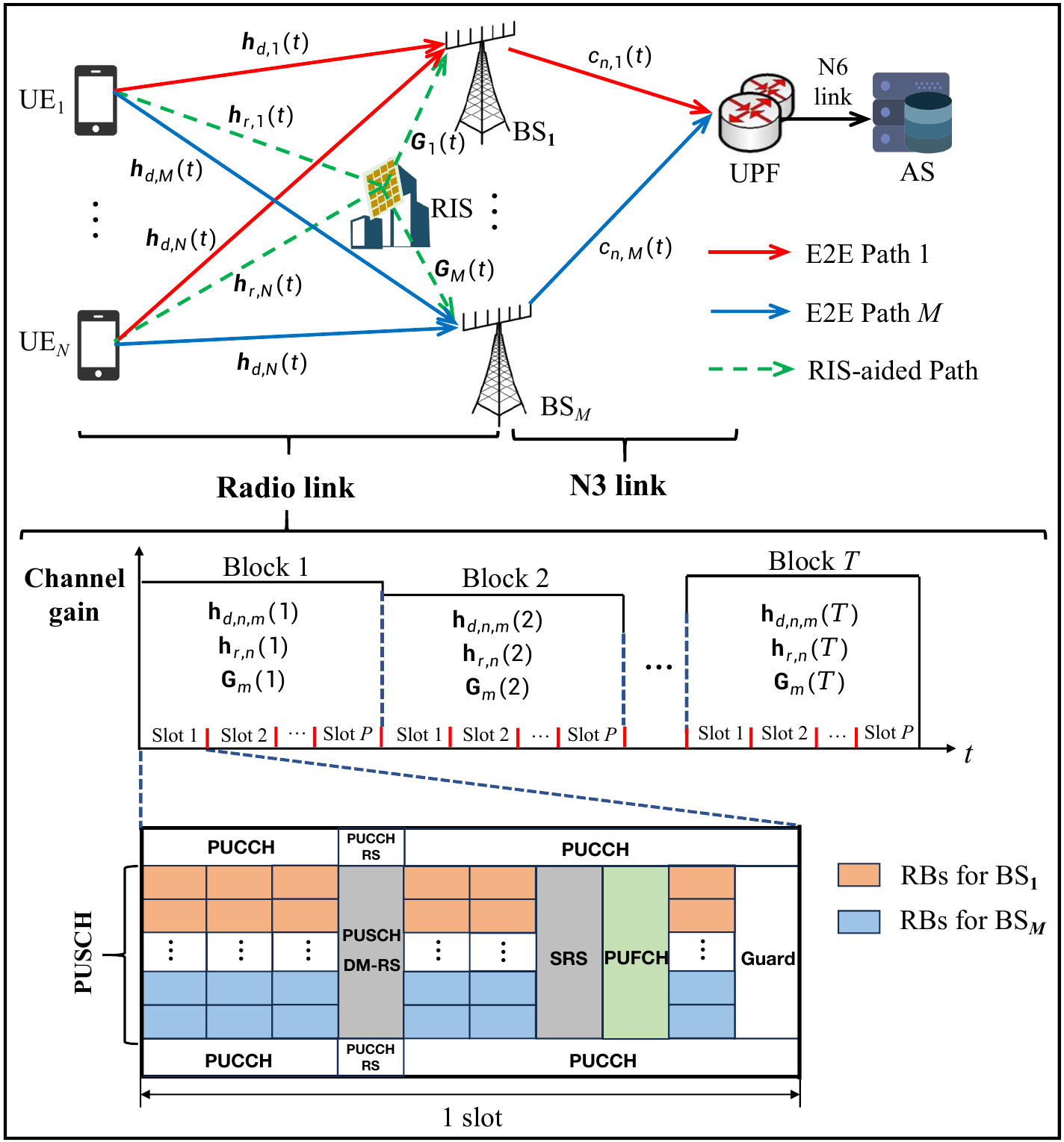}
    \caption{RIS-aided E2E Multi-path UL Transmission Architecture.}
    \vspace{-0.5cm}
    \label{fig:E2Esysarch}
\end{figure}

As Fig.~1 shows, we consider an E2E multi-path UL transmission network assisted by an Uniform Planar Array (UPA)-based RIS with $K_x \times K_y$ passive reflecting elements. 
The system serves $\mathcal{N}=\{1,\ldots,N \}$ user equipments (UEs) with each having single transmit (TX) antenna, within the coverage of $\mathcal{M}=\{1,\ldots,M\}$ base stations (BSs) of which each has $L$ receive (RX) antennas.  Each BS is connected to the User Plane Function (UPF) via
an N3 link. Then the UPF is further connected to the Application Server (AS) via an N6 link. Meanwhile, each UE transits $\mathcal{Q}=\{1,\ldots,Q\}$ UL traffic types from the user side to the UPF side, and the traffic of each type is split over the $M$ E2E paths shown in Fig.1. To minimize the E2E latency that is the sum of the latency over the radio link and the N3 link based on the proposed E2E Multi-path UL transmission architecture, a centralized controller determines: 1) the traffic splitting ratio and the UL transmit power for each UE; 2) the phase-shift for the RIS; and 3) the receive beamforming vectors for each BS.





We consider block-fading channels where the Direct UE–BS UL Channel $\mathbf{h}_{d,n,m}(t)$, UE–RIS UL Channel $\mathbf{h}_{r,n}(t)$, and RIS–BS Channel $\mathbf{G}_m(t)$ between the $n$th ($n\in \mathcal{N}$) UE and the $m$th ($m\in \mathcal{M}$) BS  in the $t$th ($t\in \mathcal{T}$) block are assumed to be constant within a block and vary independently over the $\mathcal{T} = \{1,\ldots,T\} $ blocks. Each block consists of $P$ consecutive slots with each duration being determined by the 6G numerology.  Accordingly, the joint decisions in terms of traffic splitting ratio, UL transmit power, and RIS phase shift are made at the beginning of each block and kept fixed for the whole block. Within each slot, the resource Blocks in the Pysical UL Shared Channel (PUSCH) are equally allocated to multiple BSs for the UL traffic, while the decision information as part of the control information is transmitted in the Pysical UL Control Channel (PUCCH). All the relevant feedback to UEs including Channel State Information (CSI) and the decision information (before UEs’ UL traffic transmission) is transmitted in the Pysical UL Feedback Channel (PUFCH). Note that as all UEs are within the common coverage of the $M$ BSs, each UE will reuse the same RBs allocated for each BS, thereby incurring the inter-user interferences in the UL transmission. 

\section{System Model}
\label{sec:sys_model}

In this section, we formulate a problem formulation with the algorithm solution for the proposed RIS-aided E2E multi-path framework. The UL channel consists of a direct component and an RIS-assisted cascaded component. Let $\mathbf{h}_{d,n,m}(t)\in\mathbb{C}^{L\times 1}$ denote the direct UL channel from UE$_n$ to BS$_m$, which is given by
\begin{equation}
\mathbf{h}_{d,n,m}(t)
=
\sqrt{\beta_{d,n,m}}
\left(
\eta_{n,m}\mathbf{h}^{\mathrm{LOS}}_{n,m}
+
\bar{\eta}_{n,m}\mathbf{h}^{\mathrm{NLOS}}_{n,m}
\right),
\label{eq:h_d_rician}
\end{equation}
where $\eta_{n,m}=\sqrt{\frac{\kappa_{n,m}}{\kappa_{n,m}+1}}$ and
$\bar{\eta}_{n,m}=\sqrt{\frac{1}{\kappa_{n,m}+1}}$,
$\beta_{d,n,m}$ is the large-scale fading coefficient (pathloss and shadowing),
and $\kappa_{n,m}\ge 0$ is the Rician $K$-factor. The LOS component is modeled by the BS array response vector
\begin{equation}
\mathbf{h}^{\mathrm{LOS}}_{n,m}(t)
= e^{j\varphi_{n,m}} \, \mathbf{a}_{\mathrm{BS}}\!\left(\theta^{\mathrm{AoA}}_{n,m}\right),
\label{eq:h_d_los}
\end{equation}
where $\theta^{\mathrm{AoA}}_{n,m}$ is the AoA at $\mathrm{BS}_m$ and
$\mathbf{a}_{\mathrm{BS}}(\cdot)\in\mathbb{C}^{L\times 1}$ is the BS array response.
For an $L$-element ULA, it is given by
\begin{equation}
\mathbf{a}_{\mathrm{BS}}(\theta)
= \frac{1}{\sqrt{L}}
\left[
1,\,
e^{-j\frac{2\pi d}{\lambda}\sin\theta},\,
\ldots,\,
e^{-j\frac{2\pi d}{\lambda}(L-1)\sin\theta}
\right]^{\mathsf{T}},
\label{eq:a_bs_ula}
\end{equation}
where $d$ is the inter-element spacing and $\lambda$ is the wavelength. The NLOS component accounts for rich scattering and spatial correlation
\begin{equation}
\mathbf{h}^{\mathrm{NLOS}}_{n,m}
=
\left(\mathbf{R}^{\mathrm{BS}}_{n,m}\right)^{\frac{1}{2}}
\mathbf{g}_{n,m},
\label{eq:h_d_nlos}
\end{equation}
where $\mathbf{g}_{n,m}\sim\mathcal{CN}(\mathbf{0},\mathbf{I}_L)$ and
$\mathbf{R}^{\mathrm{BS}}_{n,m}\in\mathbb{C}^{L\times L}$ is the receive correlation matrix at BS$_m$.

Let $\mathbf{h}_{r,n}(t)\in\mathbb{C}^{K\times 1}$ denote the UL channel from UE$_n$ to the RIS. A geometry-based finite-path model is adopted as
\begin{equation}
\mathbf{h}_{r,n}(t)=\sqrt{\beta_{r,n}}\sum_{s=1}^{S_r}\xi_{n,s}\,
\mathbf{a}_{\mathrm{RIS}}\!\left(\varphi^{\mathrm{AoA}}_{n,s},\theta^{\mathrm{AoA}}_{n,s}\right),
\label{eq:hrn_upa}
\end{equation}
where $\beta_{r,n}$ is the large-scale fading on the UE--RIS link, $S_r$ is the number of resolvable paths, $\xi_{n,s}$ is the complex gain of path $s$, and $\mathbf{a}_{\mathrm{RIS}}(\cdot)\in\mathbb{C}^{K\times 1}$ is the RIS array response. For a $K_x\times K_y$ UPA, the RIS array response is given by
\begin{equation}
\mathbf{a}_{\mathrm{RIS}}(\varphi,\theta)=\mathbf{a}_{x}(\varphi,\theta)\otimes \mathbf{a}_{y}(\varphi,\theta),
\end{equation}
with
\begin{equation}
\small
\mathbf{a}_{x}(\varphi,\theta)
=\frac{1}{\sqrt{K_x}}
\left[1, e^{-j\frac{2\pi d_x}{\lambda}u_x}, \ldots, e^{-j\frac{2\pi d_x}{\lambda}(K_x-1)u_x}\right]^{\mathsf{T}},
\end{equation}
\begin{equation}
\small
\mathbf{a}_{y}(\varphi,\theta)
=\frac{1}{\sqrt{K_y}}
\left[1, e^{-j\frac{2\pi d_y}{\lambda}u_y}, \ldots, e^{-j\frac{2\pi d_y}{\lambda}(K_y-1)u_y}\right]^{\mathsf{T}},
\end{equation}
where $u_x=\sin\theta\cos\varphi$ and $u_y=\sin\theta\sin\varphi$, $d_x$ and $d_y$ are the inter-element spacings along the $x$- and $y$-axes of the RIS, respectively.


Let $\mathbf{G}_m(t)\in\mathbb{C}^{L\times K}$ denote the channel from the RIS to BS$_m$. We adopt a finite-path geometric channel model as
\begin{equation}
\small
\mathbf{G}_m(t)=\sqrt{\beta_{G,m}}\sum_{s=1}^{S_G}\rho_{m,s}\,
\mathbf{a}_{\mathrm{BS}}\!\left(\vartheta^{\mathrm{AoA}}_{m,s}\right)\,
\mathbf{a}_{\mathrm{RIS}}^{H}\!\left(\varphi^{\mathrm{AoD}}_{m,s},\theta^{\mathrm{AoD}}_{m,s}\right),
\label{eq:Gm_upa}
\end{equation}
where $\beta_{G,m}$ is the large-scale fading on the RIS--BS link, $S_G$ is the number of paths, $\rho_{m,s}$ is the complex gain, and $\vartheta^{\mathrm{AoA}}_{m,s}$ is the AoA at BS$_m$. $\varphi^{\mathrm{AoD}}_{m,s}$ and $\theta^{\mathrm{AoD}}_{m,s}$ are the pitch and azimuth AoD from the RIS, respectively.


The RIS is modeled by a diagonal phase-shift matrix as
\begin{equation}
\boldsymbol{\Phi}(t)=\mathrm{diag}\big(\phi_1(t),\ldots,\phi_K(t)\big),
\label{eq:Phi_def}
\end{equation}
where $\phi_k(t)=e^{j\theta_k(t)}$ denotes the reflection coefficient of the $k$-th element. Since the RIS is passive, it satisfies the unit-modulus constraint $|\phi_k(t)|=1$, $\forall k$. Accordingly, the RIS-aided effective UL channel from UE$_n$ to BS$_m$ is
\begin{equation}
\mathbf{h}_{n,m}(t)
=\mathbf{h}_{d,n,m}(t)+\mathbf{G}_{m}(t)\boldsymbol{\Phi}(t)\mathbf{h}_{r,n}(t).
\label{eq:effective_channel}
\end{equation}

Let $x_{n,m}(t)$ denote the unit-power UL stream from UE$_n$ intended for BS$_m$ with $\mathbb{E}[|x_{n,m}(t)|^2]=1$, and let $p_{n,m}(t)\ge 0$ be the corresponding transmit power allocated to link $(n,m)$. Then the received signal at BS$_m$ is
\begin{equation}
\mathbf{y}_{m}(t)=\sum_{n=1}^{N}\sqrt{p_{n,m}(t)}\,\mathbf{h}_{n,m}(t)\,x_{n,m}(t)+\mathbf{z}_{m}(t),
\label{eq:rx_signal}
\end{equation}
where $\mathbf{z}_{m}(t)\sim\mathcal{CN}(\mathbf{0},\sigma^2\mathbf{I}_L)$ is the receiver noise. Since each UE transmits the sub-flow destined for BS$_m$ on orthogonal time--frequency resources associated with that BS, as is shown in Fig.~1, the symbol $x_{n,m}(t)$ is only observed on BS$_m$'s resource, and there is no intra-UE cross-path interference.

For each BS$_m$, define the receive combining matrix
$\mathbf{W}_{m}(t) \triangleq \big[\mathbf{w}_{m,1}(t),\ldots,\mathbf{w}_{m,N}(t)\big] \in \mathbb{C}^{L\times N}$.
Let $\mathbf{W}(t) \triangleq \{\mathbf{W}_{m}(t)\}_{m\in\mathcal{M}}$ denote the collection of all receive combiners. In this work, $\mathbf{W}(t)$ is updated using the Minimum Mean Square Error (MMSE) receive combiner based on the current $\alpha_{n,m,q}(t)$, $p_{n,m}(t)$, and $\boldsymbol{\Phi}(t)$. Particularly, BS$_m$ uses $\mathbf{w}_{m,n}(t)$ to decode UE$_n$'s uplink stream as follows:

\begin{equation}
\hat{x}_{n,m}(t)=\mathbf{w}_{m,n}^H(t)\mathbf{y}_{m}(t).
\end{equation}

The SINR at the BS$_m$ from the UE$_n$  is
\begin{equation}
\small
\gamma_{n,m}(t)=
\frac{
p_{n,m}(t)\left|\mathbf{w}_{m,n}^H(t)\mathbf{h}_{n,m}(t)\right|^2
}{
\sum_{j\ne n} p_{j,m}(t)\left|\mathbf{w}_{m,n}^H(t)\mathbf{h}_{j,m}(t)\right|^2
+\sigma^2\|\mathbf{w}_{m,n}(t)\|^2
},
\label{eq:sinr}
\end{equation}

Accordingly, the achievable UL rate (bits/second) from the UE$_n$
to the BS$_m$ is
{%
\setlength{\abovedisplayskip}{4pt}
\setlength{\belowdisplayskip}{4pt}
\setlength{\abovedisplayshortskip}{2pt}
\setlength{\belowdisplayshortskip}{2pt}
\begin{equation}
\label{eq:rate}
R_{n,m}(t)=B_m \log_2\!\big(1+\gamma_{n,m}(t)\big).
\end{equation}
}%
\noindent where $B_m$ denotes the bandwidth of the UL resource associated with BS$_m$. According to Fig.~1, the latency on each E2E path includes the latency over the radio link and the N3 link. Suppose the number of arrived packets of traffic type $q$ at UE$_n$ in block $t$ is a random variable $\lambda_{n,q}(t)$ following a Poisson distribution with mean
$\mathbb{E}\!\left[\lambda_{n,q}(t)\right]=\bar{\lambda}_{n,q}$.
Each packet of traffic type $q$ has a fixed size of $M_q$ bits, and all packets with the same type share the same size.
Meanwhile, let $n_{n,q}(t)$ denote the number of queueing packets of type $q$ at UE$_n$ at the beginning of block $t$, which also follows a Poisson distribution with mean
$\mathbb{E}\!\left[n_{n,q}(t)\right]=\bar{n}_{n,q}$.
We denote the fraction of type $q$ UL traffic from UE$_n$ routed to BS$_m$ is $\alpha_{n,m,q}(t)$, satisfying $0\le \alpha_{n,m,q}(t)\le 1$ and $\sum_{m=1}^{M}\alpha_{n,m,q}(t)=1,\ \forall n\in\mathcal{N},\ \forall q\in\mathcal{Q}$. Therefore, the corresponding propagation latency for traffic type $q$ over the radio link from UE$_n$ to BS$_m$ is obtained by 
{%
\setlength{\abovedisplayskip}{4pt}
\setlength{\belowdisplayskip}{4pt}
\setlength{\abovedisplayshortskip}{2pt}
\setlength{\belowdisplayshortskip}{2pt}
\begin{equation}
\tau^{\mathrm{UL}}_{n,m,q}(t)=
\frac{\alpha_{n,m,q}(t)\big(\lambda_{n,q}(t)+n_{n,q}(t)\big)M_q}{R_{n,m}(t)},
\label{eq:tau_ul}
\end{equation}
}%
 The backhaul latency for traffic type $q$ over N3 link from BS$_m$ to UPF is given by
{%
\setlength{\abovedisplayskip}{4pt}
\setlength{\belowdisplayskip}{4pt}
\setlength{\abovedisplayshortskip}{2pt}
\setlength{\belowdisplayshortskip}{2pt}
\begin{equation}
\tau^{\mathrm{BH}}_{n,m,q}(t)=
\frac{\alpha_{n,m,q}(t)\big(\lambda_{n,q}(t)+n_{n,q}(t)\big)M_q}{c_{n,m,q}(t)},
\label{eq:tau_bh}
\end{equation}
}%
where $c_{n,m,q}(t)$ follows a uniform distribution based on the
QoS flow characteristics. As a result, the estimated instant E2E latency for traffic $q\in\mathcal{Q}$ via the E2E path $m$ of the UE $n$ at the block $t$ is 
{%
\setlength{\abovedisplayskip}{4pt}
\setlength{\belowdisplayskip}{4pt}
\setlength{\abovedisplayshortskip}{2pt}
\setlength{\belowdisplayshortskip}{2pt}
\begin{equation}
u_{n,m,q}(t)=\tau^{\mathrm{UL}}_{n,m,q}(t)+\tau^{\mathrm{BH}}_{n,m,q}(t).
\label{eq:path_latency}
\end{equation}
}%

Our objective is to minimize the instant E2E UL latency per block of all UEs with all traffic types based on the proposed architecture in Fig.1. Accordingly, we formulate a joint optimization problem that minimizes the E2E latency per block in terms of UE configurations including traffic splitting and transmit power, the RIS configuration, and the BS configuration as below:
\begin{Prob}[RIS-aided E2E Multi-path UL Optimization]
\label{prob:P1}
\small
\begin{equation}
\begin{aligned}
& \hspace{-4.0em}\underset{\alpha_{n,m,q}(t),\,p_{n,m}(t),\mathbf{W}(t),\boldsymbol{\Phi}(t)}{\mathrm{argmin}} 
f=\sum_{n\in\mathcal{N}}\sum_{q\in\mathcal{Q}}\max_{m\in\mathcal{M}} \ u_{n,m,q}(t)  \\
\mathrm{s.t.} ~~~~&
\mathrm{C1}:~ 0 \le \alpha_{n,m,q}(t) \le 1, n\in\mathcal{N},  m\in\mathcal{M}, q\in\mathcal{Q},\\
&\mathrm{C2}:~ \sum_{m=1}^{M}\alpha_{n,m,q}(t)=1,   n\in\mathcal{N}, q\in\mathcal{Q},\\
&\mathrm{C3}:~ 0\le p_{n,m}(t)\le P_n^{tot},  n\in\mathcal{N},  m\in\mathcal{M},\\
& \mathrm{C4}:~ \sum_{m=1}^{M}p_{n,m}(t) = P_n^{tot},  n\in\mathcal{N},\\
& \mathrm{C5}:~ \|\mathbf{w}_{m,n}(t)\|^2 \le 1,  m\in\mathcal{M}, n\in\mathcal{N},\\
& \mathrm{C6}:~ \boldsymbol{\Phi}(t)=\mathrm{diag}\!\big(e^{j\theta_1(t)},\ldots,e^{j\theta_k(t)}\big),\\
& \ |\phi_k(t)|=1,  \forall k\in\{1,\ldots,K\},\\
&\mathrm{C7}: 0\leq u_{n,m,q}(t) \leq T_q^{max}, q\in\mathcal{Q}, \\
\label{eq:prob1}
\end{aligned}
\end{equation}
\vspace{-0.8cm}
\end{Prob}
\noindent where $P_n^{tot}$ is the total UL transmit power for UE${_n}$. $T_q^{\max}$ is the latency budget of traffic type $q$.



\setlength{\intextsep}{5pt}  

\begin{algorithm}[h]
\SetAlgoSkip{}
\small
\caption{Alternating Optimization (AO) for Problem 1}
\label{alg:ao_sca_compact}
\KwIn{$R_{\max}, I_{\max}, \epsilon_{\rm AO}, \epsilon_{\rm SCA}$; }
\KwOut{$\mathbf{W}^{\star}$, $\alpha^{\star}$, $p^{\star}$, $\boldsymbol{\Phi}^{\star}$;}

\textbf{Init:} Feasible $(\alpha^{0},p^{0},\boldsymbol{\Phi}^{0})$; $\mathbf{W}^{0}\leftarrow \textsc{MMSE}(p^{0},\boldsymbol{\Phi}^{0})$;\\ 
\hspace*{2.2em}$f^{0}\leftarrow f(\alpha^{0},p^{0},\mathbf{W}^{0},\boldsymbol{\Phi}^{0})$\ \\
\For{$r=0$ \KwTo $R_{\max}-1$}{
    $\mathbf W^{r+1}\leftarrow \textsc{MMSE}( p^{r},\boldsymbol\Phi^{r})$;\\
\For{$i=0$ \KwTo $I_{\max}-1$}{
$(\alpha^{i+1}, p^{i+1})\leftarrow$ SCA$(\mathbf W^{r+1},\boldsymbol\Phi^{r},\alpha^{i}, p^{i})$;\\
\If{$\frac{\left\lVert \alpha^{i+1}-\alpha^{i} \right\rVert_{2}
      +\left\lVert p^{i+1}-p^{i} \right\rVert_{2}}
{\max\!\left\{1,\ \left\lVert \alpha^{i} \right\rVert_{2}
               +\left\lVert p^{i} \right\rVert_{2}\right\}}
\le \epsilon_{\mathrm{SCA}} $}{
\textbf{break};
}
}
$(\alpha^{r+1}, p^{r+1})\leftarrow (\alpha^{i+1},p^{i+1})$;\\

        $\mathbf{V}^{r+1} \leftarrow \textsc{SDR}(\mathbf{W}^{r+1},{\alpha}^{r+1},{p}^{r+1})$;\\ 
        $\tilde{\mathbf{v}}^{r+1} \leftarrow$ Gaussian-Randomization $(\mathbf{V}^{r+1})$;
        \\
        $\mathbf{v}^{r+1} \leftarrow \tilde{\mathbf{v}}^{r+1}\
        \leftarrow $ Entry-wise projection;
    $\boldsymbol\Phi^{r+1}\leftarrow \mathrm{diag}\!\big(e^{j\angle \mathbf v^{r+1}}\big)$$\leftarrow$Unit-modulus projection\;
    
    $f^{r+1}\leftarrow f(\alpha^{r+1},p^{r+1},\mathbf w^{r+1},\boldsymbol\Phi^{r+1})$ \;
    \If{$\frac{|f^{r+1}-f^{r}|}{\max\{1,|f^{r}|\}}\le \epsilon_{\rm AO}$}{\textbf{break}\;}
}
\Return{$\{\mathbf w^{r+1},\alpha^{r+1},p^{r+1},\boldsymbol\Phi^{r+1}\}$}\;
\end{algorithm}



\setlength{\textfloatsep}{2pt}
\setlength{\floatsep}{2pt}

\begin{remark}
Problem~\ref{prob:P1} is non-convex due to the coupled SINR and UL rate expressions in Eq. \eqref{eq:sinr}--\eqref{eq:rate} and the unit-modulus RIS constraint in C6. Specifically, the SINR in Eq. \eqref{eq:sinr} couples the receive combining vectors $\mathbf{W}(t)$, the RIS phase-shift matrix $\boldsymbol{\Phi}(t)$, and the UL transmit power $\{p_{n,m}(t)\}$ through multiplicative terms in both the desired signal and interference components. Meanwhile, the achievable UL rate in Eq. \eqref{eq:rate} introduces a nonlinear mapping from SINR to the UL rate, and the E2E latency further contains inverse-rate type terms (e.g., traffic amount divided by achievable rate) that strengthens the non-convexity. Finally, the constraint $|\phi_k(t)|=1$ in C6 renders the feasible set of RIS coefficients non-convex. 
\end{remark}
\vspace{-7pt}

Motivated by Remark~1, we propose an alternating optimization (AO) framework that iteratively updates $\mathbf{W}(t)$, $\alpha(t),p(t)$, and $\boldsymbol{\Phi}(t)$ to solve \mbox{Problem~\ref{prob:P1}}. For notational convenience, we omit the index of block $t$, UE$_n$, BS$_m$, traffic type $q$ and unless otherwise stated. Specifically, $\mathbf{W}$ is updated by the MMSE receive combiner, $(\alpha,p)$ is updated by solving an Successive Convex Approximation (SCA)-based convex subproblem with at most $I_{\max}$ inner iterations (tolerance $\epsilon_{\rm SCA}$), and $\boldsymbol{\Phi}$ is updated via the Semidefinite Program (SDP) optimization. The Semidefinite Relaxation (SDR) step leads to a SDP, which can be efficiently solved by standard convex solvers. The AO iterations stop when the objective improvement between two successive iterations is below a prescribed threshold $\epsilon_{\rm AO}$, or when the maximum number of AO iterations $R_{\max}$ is reached. The overall procedure is summarized in Algorithm~\ref{alg:ao_sca_compact}\cite{scutari2018parallel,wu2021intelligent,wu2019intelligent}.
As a result, the average E2E UL latency over all $T$ blocks per UE shown in Fig.1 is expressed as
{%
\setlength{\abovedisplayskip}{4pt}
\setlength{\belowdisplayskip}{4pt}
\setlength{\abovedisplayshortskip}{2pt}
\setlength{\belowdisplayshortskip}{2pt}
\begin{equation}
\bar{U}= \frac{1}{NT}\sum_{t\in\mathcal{T}}f(\alpha^{\star},p^{\star},\mathbf{W}^{\star},\boldsymbol{\Phi}^{\star},t),
\label{eq:obj_nonconvex}
\end{equation}
}%
\noindent where $T$ should be large enough to estimate the converged one.

\section{Simulation}
\label{sec:sim}
In this section, simulations are conducted in Matlab with CVX Toolbox to quantify the performance gain of the proposed E2E multi-path architecture with algorithm. We consider a scenario of 3 BSs and 3 UEs, where the 3 BSs are placed in a triangular layout, forming spatially diverse access paths for the UEs, and each UE is within the coverage of the 3 BSs thereby the traffic can be split among the 3 E2E paths. The RIS is implemented as a 10×10 UPA with half-wavelength spacing, and the RIS-assisted paths are modeled using a Rician fading channel with $K$ factor equal to 10. Although the RIS is able to assist multiple E2E paths for each UE simultaneously, the achievable gain is heterogeneous across different paths due to distance-dependent propagation and the common use of phase-shift configuration within each block. To facilitate a clear ablation study of RIS assistance more distinguishable, the RIS is deployed close to BS 1. In this deployment, the RIS phase shifts are configured to provide constructive signal combining for the overall network rather than being dedicated to a specific BS-UE pair. As a result of the geometric proximity between the RIS and BS 1, Path~1 naturally experiences more pronounced RIS-assisted gains for the BS 1-UE 1 radio link compared with other BSs. Meanwhile, UE 2 and UE 3 are placed closer to BS 2 and BS 3, respectively, resulting in different dominant access paths. 

\begin{table}[h]
\centering
\caption{Main Simulation Parameters.}
\vspace{-0.2cm}
\label{tab:sim_para_3bs3ue}
\scriptsize
\setlength{\tabcolsep}{4.0pt}
\renewcommand{\arraystretch}{0.8}
\begin{tabular}{|c|c|c|c|}
\hline
Parameter & Value & Parameter & Value \\ \hline

Traffic types $q$
& $\{1,2\}$
& \shortstack[l]{Packet size\\ $M_q$ (bytes)}
& \shortstack[l]{T1: 10000 \\ T2: 20000} \\ \hline

\shortstack[l]{Arrival rate\\ $\lambda_{n,q}$ (pkts/s)}
& \shortstack[l]{T1: 50\\ T2: 10}
& \shortstack[l]{Latency budget\\ $T_q^{\max}$ (s)}
& \shortstack[l]{T1: 0.9\\ T2: 1.0} \\ \hline

\shortstack[l]{Noise PSD\\ $N_0$ (dBm/Hz)}
& $-174$
& \shortstack[l]{Carrier frequency\\ $f_c$ (GHz)}
& $2.6$ \\ \hline

\shortstack[l]{Total bandwidth\\ $B_{\rm tot}$ (MHz)}
& 100
& \shortstack[l]{Uplink transmit power\\ $P_{\rm tot}$ (dBm)}
& 23 \\ \hline

Pathloss model
& UMa NLoS \cite{3gpp.38.901}
& \shortstack[l]{UE speed(m/s)}
& 1 \\ \hline

\shortstack[l]{Block duration \\ $t_s$ (ms)}
& 50
& \shortstack[l]{Simulation time\\ $T_{\rm sim}$ (s)}
& 500 \\ \hline

\shortstack[l]{GBR Path 1\\ $Q_1$ (Mb/s)}
& \shortstack[l]{T1: [100,150]\\ T2: [180,220]}
& \shortstack[l]{GBR Path 2\\ $Q_2$ (Mb/s)}
& \shortstack[l]{T1: [110,140]\\ T2: [190,200]} \\ \hline

\shortstack[l]{GBR Path 3\\ $Q_3$ (Mb/s)}
& \shortstack[l]{ T1: [105,130]\\ T2: [170,210]}
& \shortstack[l]{RIS location (m)}
& (50,86.6,20) \\ \hline

\shortstack[l]{BS location(m)}
& \multicolumn{3}{c|}{(0,0,25); (433,0,25); (216.5,375,25)} \\ \hline

\shortstack[l]{UE location (m)}
& \multicolumn{3}{c|}{(151.6,93.8,1.8); (368.1,56.3,1.8); (66.5,318.8,1.8)} \\ \hline

CPU
& \multicolumn{3}{c|}{Intel Core i7 @ 2.6 GHz} \\ \hline

\end{tabular}
\end{table}

The main simulation setup is summarized in Table \ref{tab:sim_para_3bs3ue}, where the two traffic types have different latency budget characteristics. We further propose three baselines for comparison:
\begin{itemize}
 \item \textbf{Single-path (SP)}: Each UE transmits all traffic through a
 fixed access path for all traffic in all blocks.

 \item \textbf{Path-selection (PS)}: Each UE dynamically selects a better single E2E path in each block and transmits all traffic through the selected path.

 \item \textbf{Multi-path (MP) without RIS}: Each UE optimally splits its
 traffic across all available E2E paths using the approach in prior work \cite{cao2025latency}, without the RIS assistance.

\end{itemize}


\begin{figure*}[t]
\centering
\begin{minipage}{0.3\linewidth}
\centering
\includegraphics[width=\linewidth]{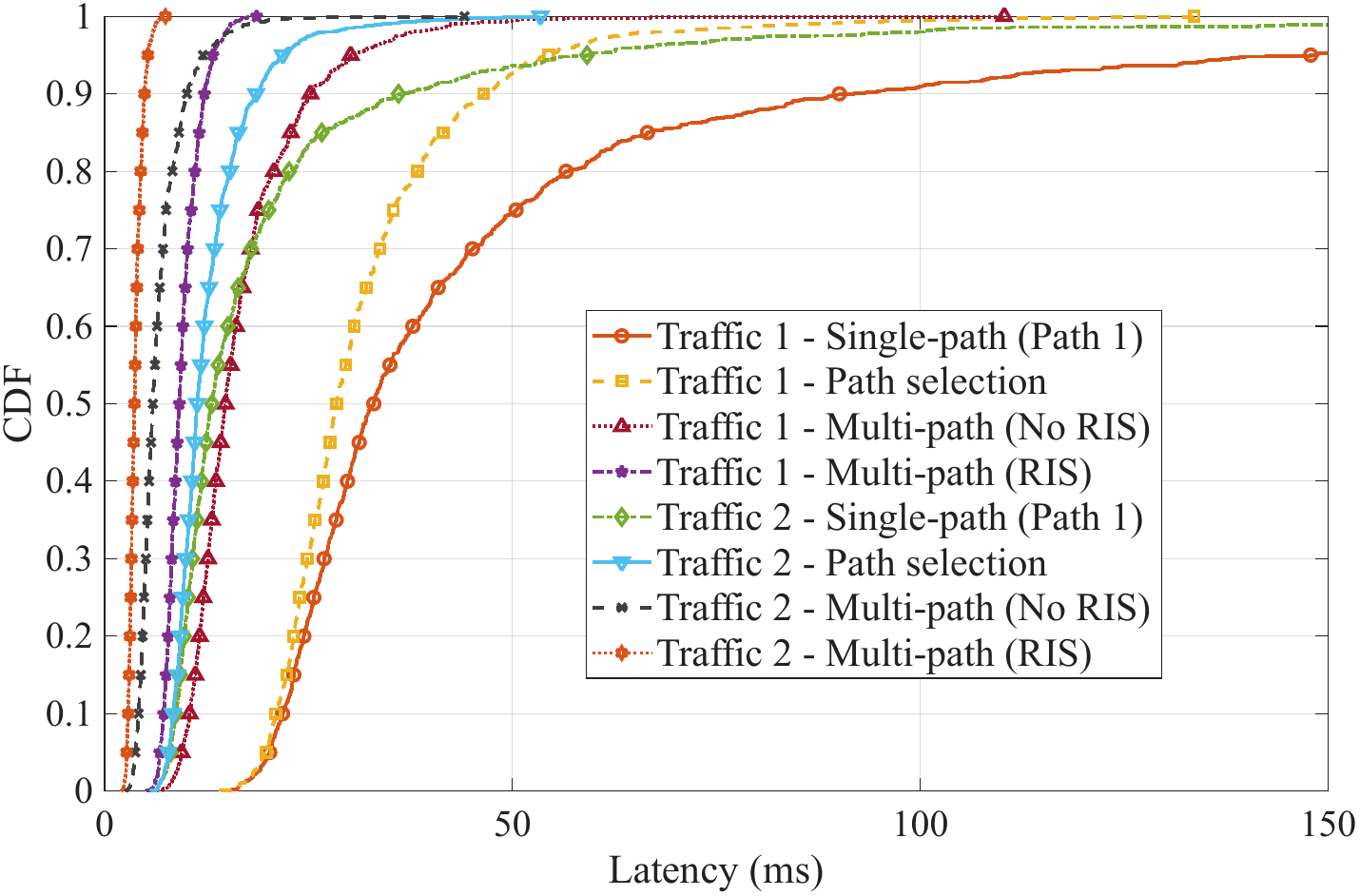}
\small (a) UE 1.
\end{minipage}
\hfill
\begin{minipage}{0.3\linewidth}
\centering
\includegraphics[width=\linewidth]{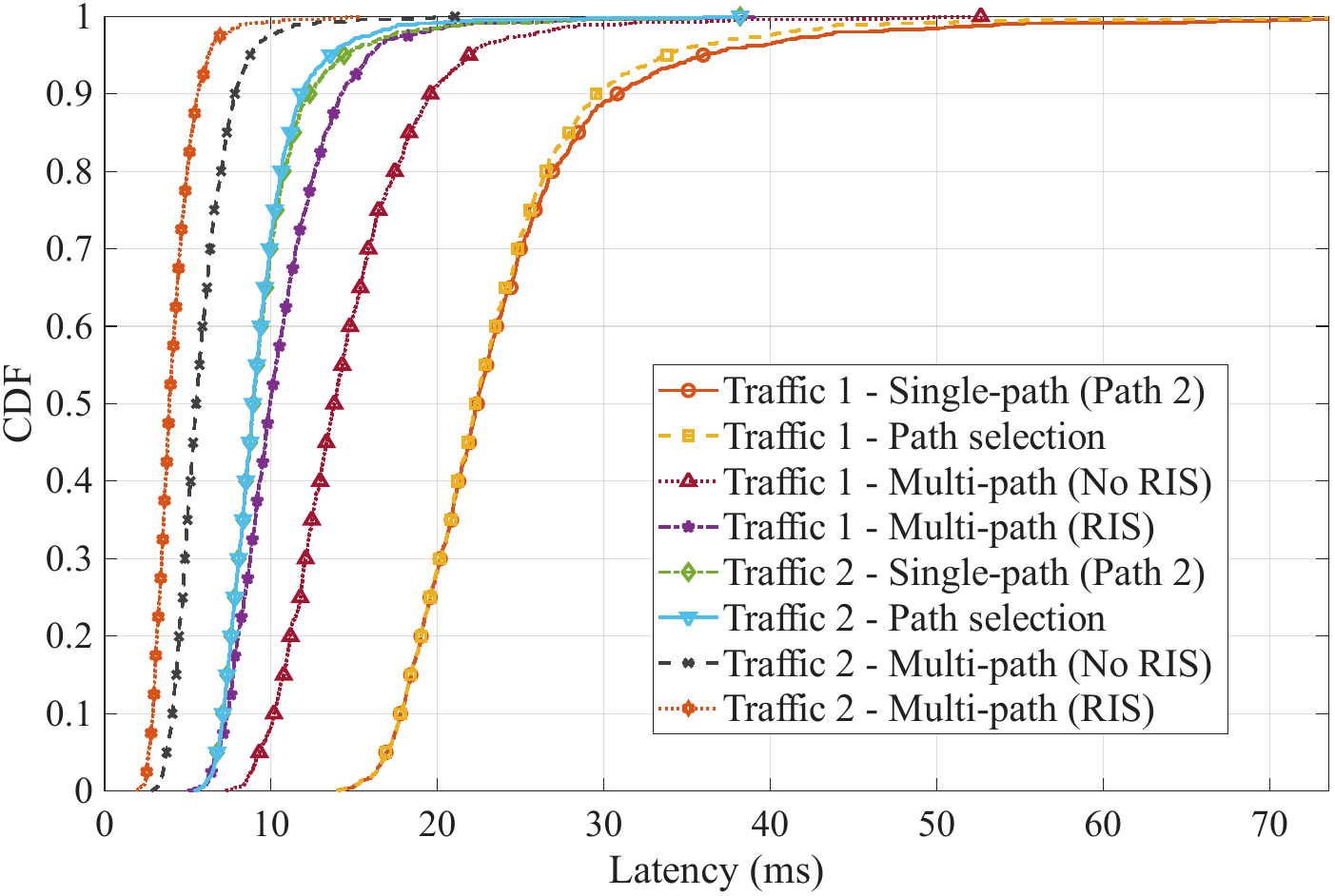}
\small (b) UE 2. 
\end{minipage}
\hfill
\begin{minipage}{0.3\linewidth}
\centering
\includegraphics[width=\linewidth]{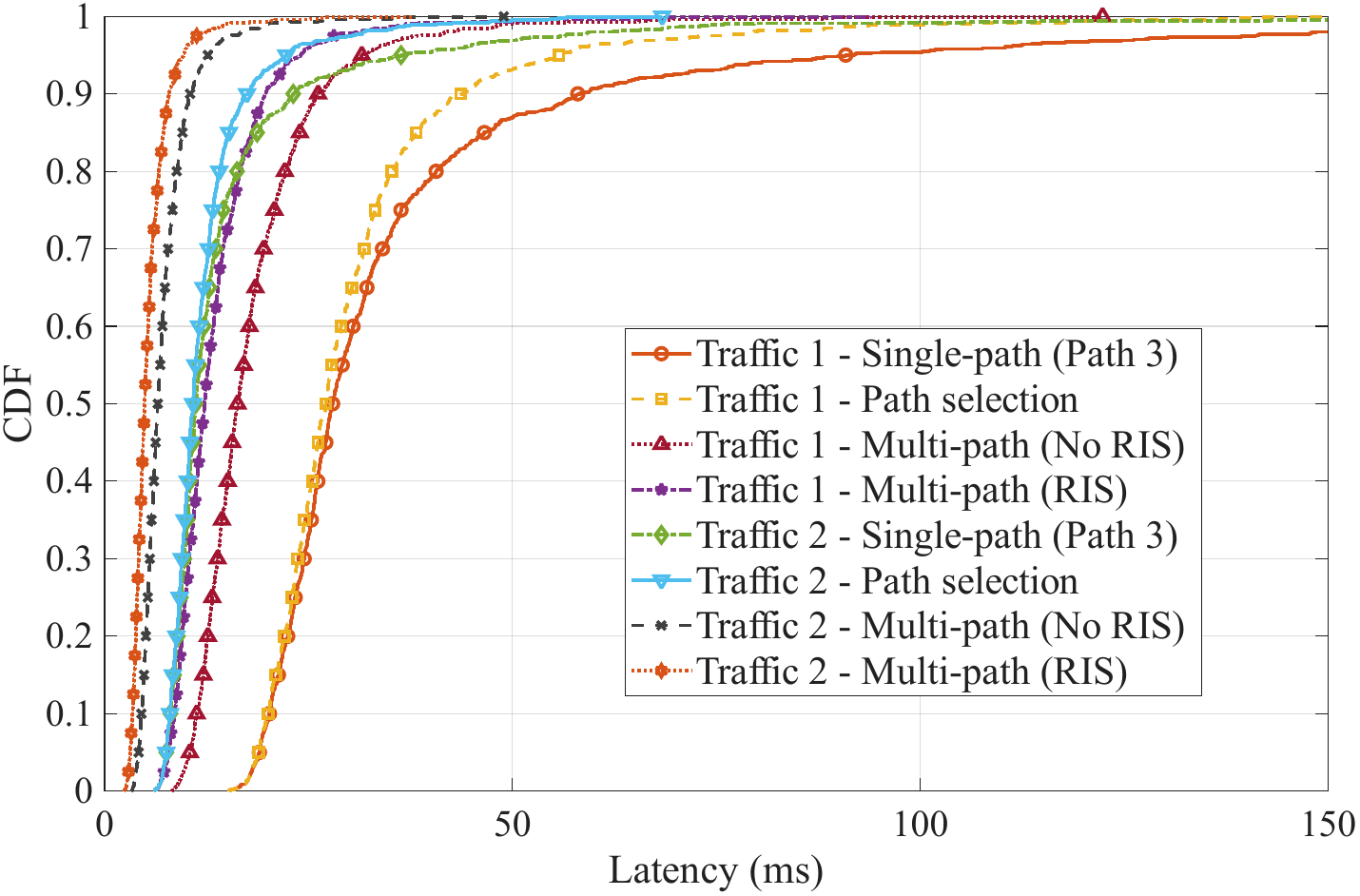}
\small (c) UE 3. 
\end{minipage}
\caption{The CDF of Instant E2E Latency per Block for Each UE.}
\label{fig:fig2}
\vspace{-0.5cm}
\end{figure*}

We first analyze the performance gain of each UE in the system. Fig.~\ref{fig:fig2} shows the cumulative distribution function (CDF) of the instant E2E latency per block for the two traffic types with different approaches. For UE 1 in Fig.~\ref{fig:fig2}(a), the RIS-assisted MP scheme consistently exhibits a left-shifted CDF, indicating the lower instant E2E multi-path latency in most blocks compared with other baselines. Specifically, the average E2E latency for UE 1 using the MP without RIS is 16.82~ms and 6.72~ms for Traffic~1 and 2, respectively, which are reduced to 9.49~ms and 3.79~ms when the RIS is enabled with latency reduction of 43.56\% and 43.58\%.  Note that the SP exhibits a pronounced long-tail behavior due to the severe radio links in some blocks. Due to the geometric proximity between the RIS, BS~1, and UE~1, the RIS-assisted multi-path scheme exhibits a more noticeable left shift in Fig.~\ref{fig:fig2}(a). In contrast, for UE~2 and UE~3 shown in Fig.~\ref{fig:fig2}(b) and (c), the latency curves of the MP with and without RIS almost overlap, indicating that the performance gain with RIS becomes weak when the RIS is too far away from the UE or the BS in the proposed system architecture.

\begin{table}[h]
\centering
\caption{Average E2E Latency (ms) per Traffic Type.}
\vspace{-0.2cm}
\label{tab:aveLatency}
\scriptsize
\setlength{\tabcolsep}{10pt}
\renewcommand{\arraystretch}{1}
\begin{tabular}{|c|c|c|c|c|}
\hline
Traffic & MP with RIS (Ours) & MP \cite{cao2025latency} & SP & PS \\
\hline
1 & \textbf{11.283}& 16.587 & 39.766 & 28.786 \\
\hline 2 & \textbf{4.494}& 6.630 & 15.896 & 11.506 \\
\hline
\end{tabular}
\end{table}

We further investigate the performance gain of the whole system based on $\bar{U}$ in Eq.(\ref{eq:obj_nonconvex}). We track the average E2E latency of traffic 1 and 2 respectively, which are summarized in Table~\ref{tab:aveLatency}. The MP achieves the lower average latency than both SP and PS for the two traffic types. Moreover, enabling the RIS for MP can further reduce the average E2E latency, confirming the effectiveness of RIS. Specifically, the MP with RIS reduces the average E2E latency by approximately 32.0\% for Traffic~1 and 32.2\% for Traffic~2, compared with the MP without RIS. In addition, compared with the SP, the MP with RIS  reduces the average E2E latency up to 64\% for both Traffic~1 and ~2. Overall, the numerical results demonstrate that while the E2E multi-path architecture is the key factor in reducing the E2E latency, the RIS is also able to primarily provide additional performance gain especially when more favorable RIS-assisted radio links are present, leading to an around 32\% performance gain in the investigated scenario.

\section{Conclusion}
\label{sec:con}
This paper proposed a RIS-assisted E2E multi-path UL transmission optimization approach to achieve the minimum average E2E latency that accounts for wireless and backhaul delays for 6G time-sensitive services. 
We formulate an optimization problem by jointly optimizing the UL traffic-splitting ratio, UL transmit power, receive combining vector, and RIS phase shift. 
An alternating optimization approach combining with the SCA and SDR approaches is then developed to solve the non-convex problem. 
Finally, the numerical results from simulation demonstrate that the proposed E2E MP architecture outperforms the baselines such as SP and PS schemes, meanwhile, the RIS aided for MP yields further gains by lowering the average E2E latency up to 43\% for a single user and 32\% for the whole system, which supports the UL transmission of 6G time-sensitive services with more stringent QoS requirements.

\bibliographystyle{IEEEtran}
\bibliography{ref}
\end{document}